\documentclass[12pt,authoryear]{elsarticle}

\usepackage{amssymb}
\usepackage{amsmath}

\usepackage{lineno}

\usepackage{booktabs}
\usepackage{threeparttable}

\journal{Transportation Research Part F: Transport Psychology}

\begin{document}

\begin{frontmatter}

\title{Which Moments Matter? Heuristics of Remembered Travel Experience in Public Transport} 

\author[DLR]{Esther Bosch\corref{cor1}}
\ead{Esther.Bosch@dlr.de}
\ead[url]{https://orcid.org/0000-0002-6525-2650}

\author[DLR]{Klas Ihme}
\ead[url]{https://orcid.org/0000-0002-7911-3512}


\author[BUM]{Stefan Bohmann}
\ead[url]{https://orcid.org/0000-0002-...}

\cortext[cor1]{Corresponding author}

\address[DLR]{German Aerospace Center (DLR), Institute of Transportation Systems,\\
Braunschweig, Germany}
\address[BUM]{University of the Bundeswehr,\\
Munich, Germany}

\begin{abstract}
Understanding how travelers form overall evaluations of public transport journeys is critical for improving travel satisfaction and encouraging sustainable mode choice. While travel satisfaction is discussed to influence attitudes and future behavior, the cognitive rules by which moment-to-moment experiences are aggregated into retrospective evaluations remain poorly understood in transport research. Drawing on psychological theories of experienced and remembered utility, this study investigates which temporal aggregation heuristics best predict post-trip travel satisfaction.

Using a smartphone-based experience sampling approach, we collected high-frequency on-trip experience ratings and post-trip evaluations for 2576 real-world public transport trips across three German cities. Travel experience was assessed every five minutes during trips using a multi-item scale, allowing direct comparison of competing aggregation rules, including mean experience, peak–end, minimum–end, final moment, and trip duration. Multilevel regression models were estimated to evaluate the explanatory power of each heuristic.

Results show that retrospective travel satisfaction is best predicted by a Minimum–End heuristic, combining the most negative moment of the journey and the final experience. Models based on mean experience, peak–end rules, final moment alone, or trip duration performed substantially worse. This pattern indicates that both negative extremes and the final phase of a journey independently contribute to remembered evaluations, rather than overall satisfaction reflecting an average of momentary experiences.

The results have important implications for theory and practice, suggesting that targeted interventions at critical negative moments and at trip endings may yield substantial improvements in remembered satisfaction and, ultimately, support shifts toward sustainable mobility.
\end{abstract}


\begin{highlights}
\item High-frequency experience sampling captures moment-to-moment travel experience during real-world public transport trips.
\item Retrospective travel satisfaction is best predicted by a Minimum–End heuristic rather than by mean experience or duration.
\item The most negative moment and the final experience disproportionately shape remembered public transport journeys.
\item Fine-grained withon-trip measurements outperform coarse or single-item approaches for testing evaluative heuristics.
\item Targeting worst moments and trip endings offers leverage points for improving public transport satisfaction and mode choice.

\end{highlights}


\end{frontmatter}


%

\section{Introduction}

A large-scale shift from private car use to public transport (PT) is a central challenge for a transition towards sustainable mobility. Achieving this shift is crucial to providing door-to-door accessibility for all travelers while reducing environmental impacts. However, despite substantial investment in PT infrastructure and service quality, many travelers remain reluctant to switch modes. Short-term interventions such as monetary incentives can temporarily encourage the use of sustainable transport modes, but their effects are typically limited to the duration of the intervention and the targeted behavior itself \citep{kaiser2020financial, steinhorst2018effects}. To achieve lasting change, behavioral interventions must address not only situational barriers but also the underlying motivations and attitudes toward public transport.


One promising pathway for strengthening positive attitudes toward PT lies in enhancing travel satisfaction. In the Travel Mode Choice Cycle described by \citet{de2022attitude}, the choice of travel mode influences the evaluation of the journey (travel satisfaction), which then shapes the mode-specific attitude and intention to use that mode in the future. Recent extensions of the model \citep{lim2024effects} explicitly incorporate within-trip subjective experiences influenced by cognitive appraisals of events such as stations, comfort, cleanliness, and availability \citep{susilo2014exploring}. Consistent with this, empirical evidence shows that service-related factors such as on-board cleanliness and comfort, courteous and helpful behavior from operators, safety, punctuality, and service frequency are particularly influential for satisfaction, while perceptions of value-for-money, on-board safety and cleanliness, and interactions with personnel are central to increasing loyalty and long-term commitment to PT \citep{van2018influences}. Furthermore, feeling safe and the frequency of PT use have been identified as important predictors of perceived accessibility \citep{lattman2016perceived}. If improving the travel experience can enhance travel satisfaction, and if satisfaction influences attitudes and subsequent choices, then targeted improvements to these specific service attributes could contribute to long-term behavioral change.

However, the mental process by which travelers integrate moment-to-moment experiences into an overall evaluation of a completed journey remains poorly understood in the PT domain. Psychological research on the evaluation of temporally extended episodes suggests that retrospective assessments are subject to systematic biases. A robust finding, known as the peak–end rule, is that global retrospective evaluations are disproportionately influenced by the most intense moment (peak) and the final moment (end) of the experience, rather than by its duration or average moment-to-moment quality \citep{kahneman1993more, redelmeier1996patients, ariely2000does}. In some contexts, such as medical procedures, small changes to the end of the experience can meaningfully alter global evaluations and subsequent behavior \citep{redelmeier2003memories}. While PT journeys differ from such medical episodes in structure and salience of duration, the possibility that remembered satisfaction is similarly dominated by salient or final moments could offer opportunities for behavioral intervention.


A related stream of work has emphasized the importance of critical incidents during travel as drivers of satisfaction and emotional well-being. Critical incidents are discrete positive or negative episodes, such as an unexpected delay, unhelpful staff interaction, or a pleasant social encounter, that disproportionately affect remembered travel experience \citep{bitner1994critical, friman2004structure, gremler2004critical}. \citet{friman2001frequency} demonstrated that negative critical incidents in public transport, such as repeated delays or unreliable information, not only shape immediate post-trip mood but also accumulate over time into stable cognitive evaluations of service quality. Theoretical models formalize overall evaluations as an aggregation of incident-specific emotional responses, weighted by memory-decay parameters and primacy–recency effects \citep{schreiber2000determinants, ariely1998combining, davelaar2005demise}. This perspective suggests that remembered travel satisfaction may depend not only on the average of momentary experiences but also on the salience and frequency of particularly impactful events.


Finally, the measurement of travel-related emotional well-being has itself been subject to increasing research interest. A comprehensive review highlights that most existing studies rely on retrospective reports of mood or satisfaction, while only a smaller number capture current mood before, during, and after travel \citep{garling2020review}. The review concludes that instantaneous, repeated assessments provide the most valid basis for aggregating commuting-related emotional well-being, yet such methods are often impractical at scale. Importantly, this literature also draws attention to the distinction between experienced utility, i.e. the subjective quality of the trip as lived, and decision utility, i.e. the expected utility that guides travel choices \citep{kahneman1997back, loewenstein2008hedonic}. Because decision utility often diverges from experienced utility, especially under conditions of uncertainty or habit, interventions that improve travelers’ experienced utility may gradually shift decision-making toward sustainable modes. 

In the literature, related but partly distinct terms are used to describe retrospective evaluations of trips: psychologists often refer to this as remembered utility or recalled happiness, whereas transport researchers typically use the construct of travel satisfaction, which operationalizes such evaluations with domain-specific scales. In practice, all three terms capture travelers’ post-hoc assessments of their journeys, though they emphasize different disciplinary perspectives. Throughout this paper, we will refer to the construct as travel satisfaction. We use the term 'travel experience' to refer to the momentary utility during a journey.

A study that also used experience sampling, however not clearly concentrated on traveling, is \citet{raveau2016smartphone}, who measured moment utility as real-time happiness in a smartphone-based study, and remembered utility as retrospective happiness. They found that retrospective evaluations of happiness show three cognitive biases: peak–end effects, hedonic adaptation toward neutrality, and duration neglect, where activity length influences momentary but not recalled happiness. Regarding moment and remembered utility in the PT context, \citet{susilo2014exploring} find that satisfaction with the primary trip stage is strongly linked to overall trip satisfaction. \citet{bosch2025travel} similarly applied experience sampling in real-world PT trips, identifying spatial hot spots of stress and cold spots of satisfaction across tram, train, and bus journeys, though without examining how momentary ratings aggregate into retrospective evaluations. On the other hand, \citet{abenoza2019does} found that duration-weighted averaging rules outperform heuristic rules such as peak–end. However, these previous studies of heuristics in remembered utility are constrained by methodological limitations: the absence of withon-trip measurement or the exclusive use of single-item satisfaction measures hinder the ability to detect the influence of specific moments or events on remembered satisfaction.

The present study addresses these gaps by collecting high-resolution travel experience data during and immediately after intermodal trips in one small and two large German cities. Using an eight-item scale, we measured travel experience during intermodal public transport trips, capturing both in-vehicle and transfer stages. This allows us to directly compare momentary evaluations with retrospective trip ratings and to test competing aggregation rules. Our findings indicate that the overall retrospective evaluation of the trip is best predicted by a minimum end rule similar to what \citet{kahneman1993more} suggested.

From a practical perspective, this has direct implications for service improvement strategies. If stakeholders can identify and mitigate the worst moments of the journey or ensure that the final part of the trip leaves a positive impression, they may achieve a disproportionately large improvement in remembered satisfaction. Because remembered satisfaction influences mode-related attitudes and, ultimately, future mode choice, targeted interventions at these critical points could support long-term behavioral change toward sustainable mobility.

\section{Methods} 

\subsection{Procedure}
Data collection took place in Hamburg, Berlin, and Tuttlingen (all in Germany) from January to December 2025. The dataset reported here partially overlaps with a previously published study [citation removed for blind review]
; that study examined a distinct research question using a subset of the journeys analysed here.

Participants were instructed to document at least six everyday trips lasting a minimum of 15 minutes each within a three-week period after enrollment using experience sampling. Across all recorded routes, the mean trip duration was 32 minutes (SD = 14), with durations ranging from 15 to 135 minutes and a median of 29 minutes. Participants could receive a maximum compensation of 70€, comprising 20€ for completing the first three trips, a 30€ bonus upon completion of all six trips, 10€ for achieving a questionnaire response rate of at least 95\%, and an additional 1€ per optional trip (up to ten trips).

\subsection{Participants}

A total of 349 participants were recruited through in-vehicle advertisements in Hamburg as well as flyers in Berlin and Tuttlingen. In total, 280 participated in Hamburg, 54 in Berlin and 15 in Tuttlingen. Of those participants in the sample willing to report these demographic data,  5 reported to identify as diverse, 158 identified as male, and 161 as female. The average age was 34 years (± 12). Among participants who provided information on their mobility habits, 6 used public transport once per week, 108 several times per week, and 214 daily. On average, participants used public transport 71 minutes per week for commuting, 31 minutes for free-time, 18 minutes for groceries, 18 to get to appointments and 19 to get to business trips.	

Prior to participation, all individuals received information regarding study procedures and data protection measures and provided written consent in line with the General Data Protection Regulation (GDPR). Ethical approval was granted by the institution’s ethics committee (reference no. 04/25).

\subsection{Materials}
After signing the informed consent form and completing a demographic survey, participants were asked to download the institute’s research app.

Before starting each trip, participants selected the route they intended to take within the app, enabling the recording of both available route options and the chosen route, including line identifiers and departure times.

They were also asked at the beginning of each journey whether they had any special requirements (e.g., accessibility needs, traveling with children, carrying luggage).

During the journey, the app delivered a pop-up questionnaire every five minutes. This survey is referred to in the paper as the Experience Sampling Questionnaire, and its psychometric proporties are described by [the same citation removed for blind review]
. Under the heading “Evaluation of journey” (translated from German), participants responded to the following items using discrete (integer-stepped) sliders:

\begin{enumerate}
\item with the current situations on my journey I am 1 (totally dissatisfied) to 5 (totally satisfied)
\item the journey at the moment is 1 (miserable) to 5 (excellent)
\item for achieving my goals, this journey is 1 (a hindrance) to 5 (favourable)
\item at the moment my journey is going 1 (not smoothly) to 5 (smoothly)
\item I currently find my journey 1 (unpleasant) to 5 (pleasant)
\item compared to an ideal trip, my trip so far is 1 (the worst) to 5 (the best)
\item the current situation on my journey is 1 (worse than expected before starting to travel) to 5 (better than expected before starting to travel)
\item Based on my previous experience with traveling this journey is 1 (below average) to 5 (above average)
\end{enumerate}

A composite score for travel experience was computed for each questionnaire by averaging the responses to these eight items.

A shortened version of the questionnaire, retaining only items 2, 3, and 8 based on their psychometric properties (manuscript under preparation), was introduced in September, resulting in 581 of 2576 post-trip questionnaires being completed in abbreviated form. Post-trip travel satisfaction scores for these cases were computed by averaging responses across the three retained items. To verify that this did not affect the pattern of results, all analyses were repeated excluding the shortened version; findings were consistent (see ~\ref{app3}).

On the same screen, participants could set tick marks to select any events that occurred during the last five minutes and influenced their travel experience. However, this data was not analyzed in this paper. 

Each answered questionnaire was saved with a timestamp and GNSS-based location.

At the end of each trip, the app automatically launched the post-trip questionnaire, which contained the following items assessing the overall evaluation of the journey:

\begin{enumerate}
\item with the situations at the end of my journey I am 1 (totally dissatisfied) to 5 (totally satisfied)
\item the journey went 1 (miserable) to 5 (excellent)
\item for achieving my goals, this journey was 1 (a hindrance) to 5 (favourable)
\item my journey was going 1 (not smoothly) to 5 (smoothly)
\item I found my journey 1 (unpleasant) to 5 (pleasant)
\item compared to an ideal trip, my trip was 1 (the worst) to 5 (the best)
\item the overall situation on my journey was 1 (worse than expected before starting to travel) to 5 (better than expected before starting to travel)
\item Based on my previous experience with traveling this journey was 1 (below average) to 5 (above average)
\end{enumerate}

These items were formulated to be as close as possible to the during-trip items but were phrased to capture the participant’s overall evaluation of the entire journey. 


\subsection{Data Processing}
Trips with fewer than three on-trip experience reports or a total duration of less than 15 minutes were excluded from the analysis. Trip duration was calculated as the time interval between the first and last on-trip questionnaire, plus an additional five minutes to account for the sampling interval. This adjustment was necessary because participants occasionally selected their intended route in the app several hours before the actual start of the journey, as indicated by the timestamp of the first on-trip report. After applying these exclusion criteria and preprocessing steps, a total of 2576 trips were retained for analysis.

\subsection{Analyses}

To determine which temporal summary statistic is most suitable for explaining post-trip experience, we estimated a series of non-nested mixed-effects models. Although each of the post-trip experience ratings were ordinal, averaging across them yields a quasi-continuous measure. Each model included a random intercept for participant to account for repeated measurements, and a single fixed-effect predictor corresponding to one temporal aggregation rule. Comparing non-nested models using Akaike’s Information Criterion (AIC) and Bayesian Information Criterion (BIC) is appropriate in this context, as it allows assessing the relative fit of competing heuristics with comparable model complexity. This approach follows prior work in travel satisfaction research, such as \citet{abenoza2019does}, who similarly estimated separate models for each aggregation rule to examine how different temporal summaries relate to overall trip evaluations. In line with this methodological tradition, we use AIC and BIC to identify which temporal summary offers the most parsimonious explanatory value for post-trip experience.


Based on \citep{kahneman1993more}, we test the aggregation rules of
\begin{enumerate}
    \item the most extreme value (i.e., with the highest deviation from the mean journey rating) plus the final value ('Peak-End')
    \item the most negative value plus the final value (Minimum-End)
    \item the mean value ('Mean')
    \item only the final value ('end')
    \item only the minimal value ('min')
    \item the duration of the whole journey
\end{enumerate}

\section{Results}

\begin{figure}[h]  
\centering
\includegraphics[width=1\linewidth]{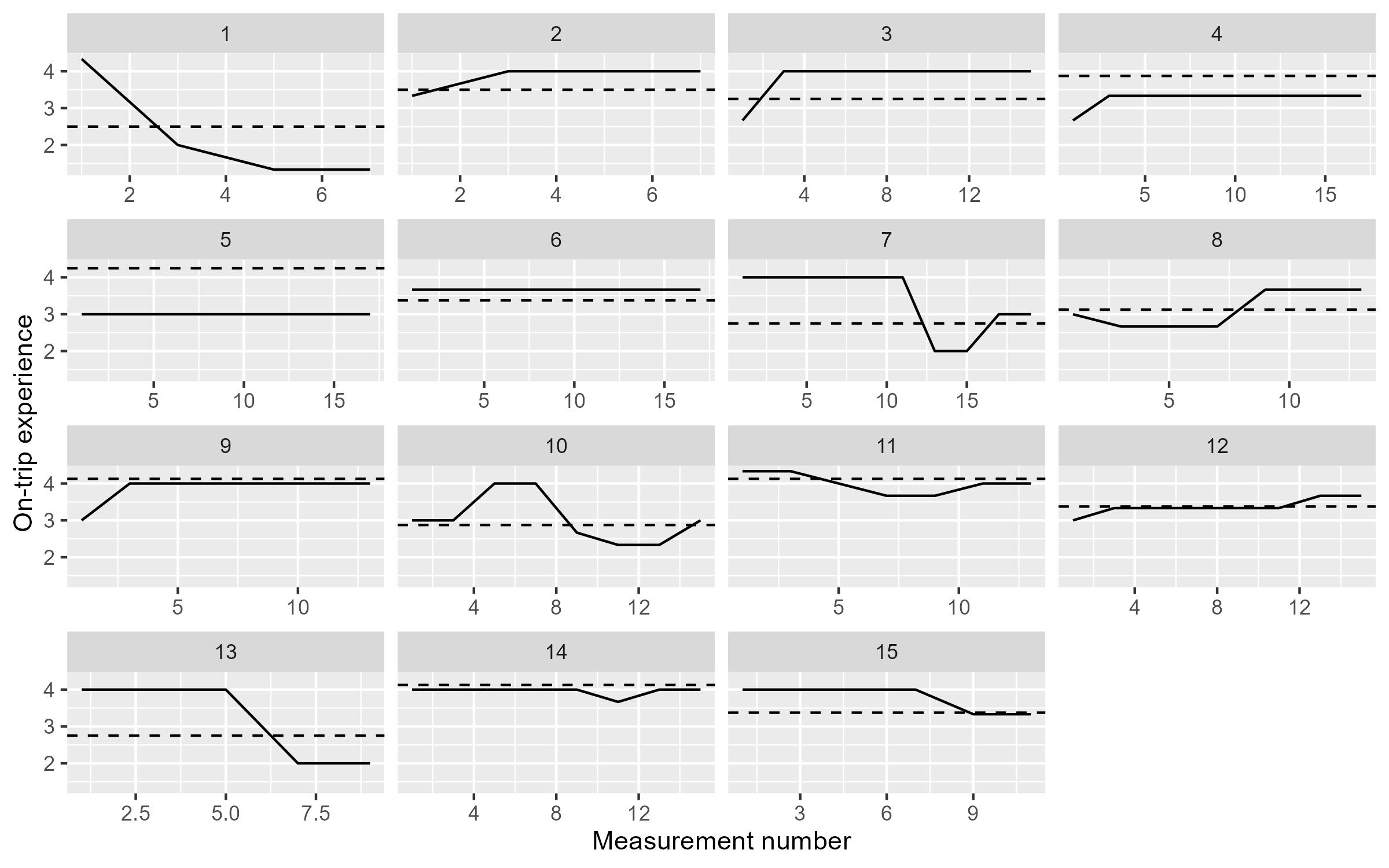}
\caption[Raw data overview]{\label{fig:exp} Exemplary data of a participant. Each subfigure is one trip. The y axis displays the ontrip travel experience rating as averaged over all items per measurement timepoint. The dashed line indicates the overall travel satisfaction with the trip.}
\end{figure}

\begin{figure}[h]  
\centering
\includegraphics[width=1\linewidth]{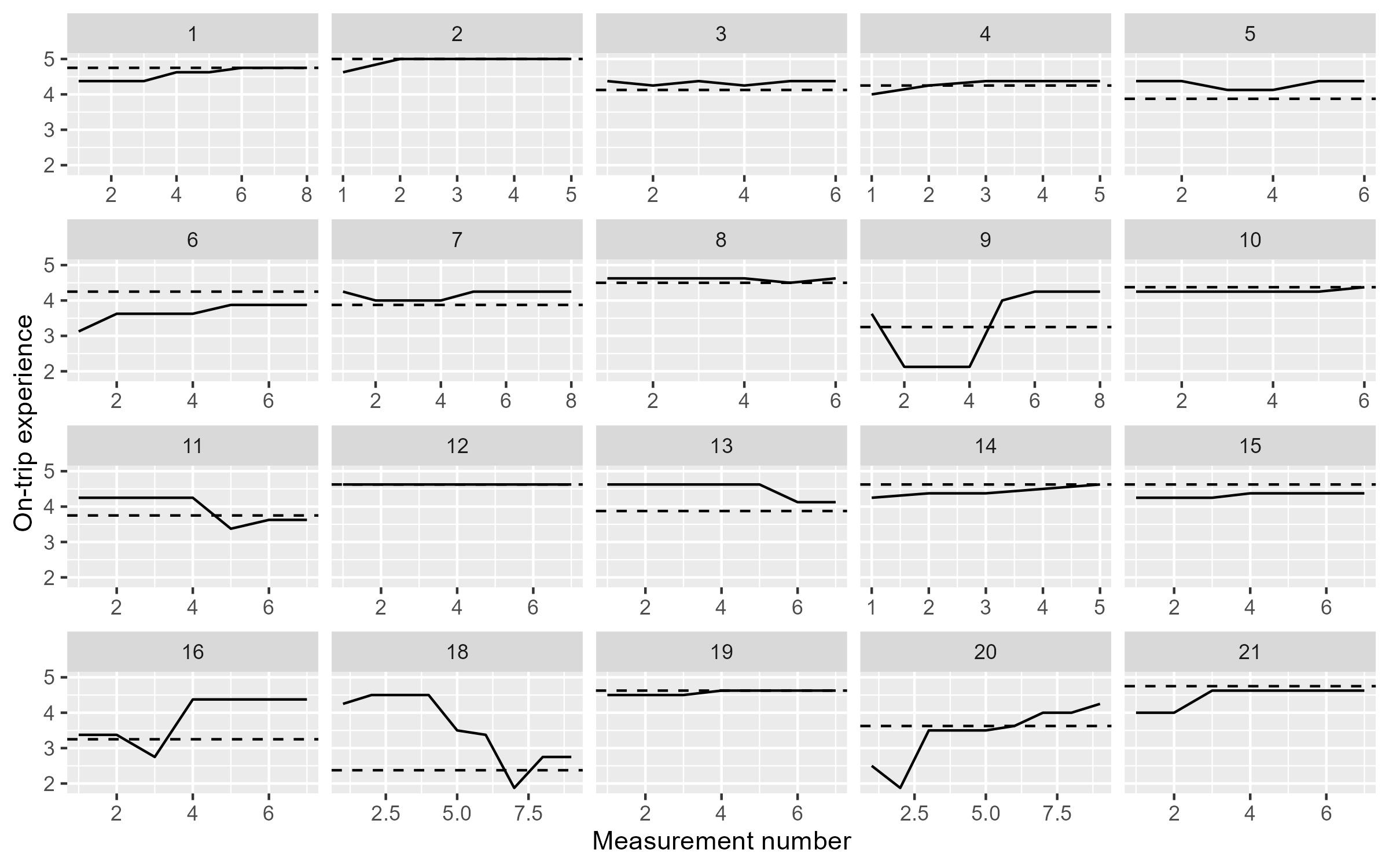}
\caption[Raw data overview]{\label{fig:exp2} Exemplary data of another participant. Each subfigure is one trip. The y axis displays the ontrip travel experience rating as averaged over all items per measurement timepoint. The dashed line indicates the overall travel satisfaction with the trip.}
\end{figure}

Exemplary data of two participants can be found in Figures \ref{fig:exp} and \ref{fig:exp2}.

Comparison of an intercept-only model with a random-intercept model accounting for participant-level variance indicated that a multilevel approach was warranted. The random-intercept model yielded substantially lower AIC (5937.9 vs. 6143.7) and BIC (5955.5 vs. 6155.4), and the intraclass correlation coefficient (ICC = .20) confirmed that 20\% of variance in post-trip experience ratings was attributable to between-participant differences. Participant was therefore included as a random effect in all subsequent models.

\begin{table}[ht]
\centering
\caption{Comparison of models predicting post-trip experience based on different heuristics}
\label{tab:model_comparison}
\vspace{0.3cm}
\begin{tabular}{lccc}
\hline
\textbf{Model} & \textbf{df} & \textbf{AIC} & \textbf{BIC} \\
\hline
Mean        & 4 & 3767.97 & 3791.39 \\
Peakend     & 4 & 3920.37 & 3943.79 \\
MinEnd      & 4 & 3329.58 & 3352.99 \\
End & 4 & 3509.11 & 3532.57 \\
Min & 4 & 3857.95 & 3881.37 \\
Duration & 4 & 5844.33 & 5867.74 \\
\hline
\end{tabular}

\begin{flushleft}
\textit{Note.} df = degrees of freedom (number of estimated model parameters: fixed intercept, fixed slope, random intercept variance, residual variance); AIC = Akaike Information Criterion; BIC = Bayesian Information Criterion. Lower AIC and BIC values indicate better model fit.
\end{flushleft}
\end{table}

The model using the Minimum–End heuristic yielded the lowest AIC (3329.58) and BIC (3352.99), indicating the most favorable trade-off between goodness of fit and model complexity. In contrast, models based on alternative temporal summaries, such as the mean experience (AIC = 3767.97), the peak–end rule (AIC = 3920.37), the final moment only (AIC = 3509.11), the minimum moment only (AIC = 3857.95), or trip duration (AIC = 5844.33) showed substantially poorer fit (see also Table \ref{tab:model_comparison}). 
Taken together, these results suggest that the combination of the most negative moment and the final experience is the most informative single predictor of post-trip experience among the aggregation rules tested.

To examine whether the observed similarity between the final on-trip rating and the post-trip satisfaction rating reflected a recency effect rather than genuine minimum-end processing, we conducted two additional analyses. First, we reasoned that participants who varied substantially in their on-trip ratings across measurement points were less likely to anchor each response on the preceding one. We therefore repeated all model comparisons restricting the sample to participants whose within-trip on-trip variance exceeded thresholds of 0.25, 0.50, 0.75, and 1.00 (see ~\ref{app1}). Results were consistent across all thresholds. Second, we replaced the final on-trip rating with the penultimate on-trip rating as a predictor. The penultimate rating was completed at most ten minutes before journey's end and was not the last questionnaire participants encountered before completing the post-trip scale, making it a plausible but weaker recency candidate. Model comparisons with this substitution yielded the same pattern of results, with the minimum-penultimate composite emerging as the strongest predictor of post-trip travel satisfaction (see ~\ref{app2}). Although we cannot entirely rule out residual recency effects, the convergence of these two analyses suggests that on-trip ratings were largely given independently of one another.

The final model predicting post-trip experience from the minimum-end composite is reported in Table~\ref{tab:mixed_model}. The minimum-end composite explained 64.9\% of variance in post-trip travel satisfaction (marginal $R^2 = .649$), rising to 70.1\% when accounting for between-participant differences (conditional $R^2 = .701$).

\begin{table}[ht]
\centering
\caption{Linear Mixed-Effects Model Predicting Post-Trip Experience from MinEnd}
\label{tab:mixed_model}
\begin{threeparttable}
\begin{tabular}{lcccccc}
\toprule
 & $b$ & $SE$ & $df$ & $t$ & $p$ \\
\midrule
\multicolumn{6}{l}{\textit{Fixed Effects}} \\
\quad Intercept & 0.636 & 0.048 & 2278 & 13.12 & < .001 \\
\quad MinEnd    & 0.869 & 0.013 & 2528 & 67.35 & < .001 \\
\midrule
\multicolumn{6}{l}{\textit{Random Effects}} \\
\quad $\tau$ (Intercept) & 0.183 & & & & \\
\quad $\sigma$ (Residual) & 0.438 & & & & \\
\midrule
\multicolumn{6}{l}{\textit{Model Fit}} \\
\quad Marginal $R^2$           & \multicolumn{5}{l}{.649} \\
\quad Conditional $R^2$        & \multicolumn{5}{l}{.701} \\
\quad Observations             & \multicolumn{5}{l}{2576} \\
\quad Participants ($N$)             & \multicolumn{5}{l}{349} \\
\bottomrule
\end{tabular}
\begin{tablenotes}
\small
\item \textit{Note.} $b$ = unstandardized coefficient; $SE$ = standard error;
$df$ = Satterthwaite-approximated degrees of freedom;
$\tau$ = standard deviation of the random intercept (between-participant);
$\sigma$ = residual standard deviation (within-participant).
Marginal $R^2$ reflects variance explained by fixed effects only;
conditional $R^2$ reflects variance explained by both fixed and random effects.
Model estimated via maximum likelihood.
\end{tablenotes}
\end{threeparttable}
\end{table}


\clearpage

\section{Discussion}

This study contributes to the growing literature on the retrospective evaluation of temporally extended experiences by providing novel empirical evidence from the domain of public transport. Building on psychological research on experienced and remembered utility (e.g., \citet{kahneman1993more, ariely1998combining, redelmeier1996patients}) and recent transport studies using experience sampling (e.g., \citet{susilo2014exploring, abenoza2019does}), this work advances understanding of how moment-to-moment travel experiences are cognitively aggregated into an overall evaluation of a completed journey. By combining high-frequency on-trip assessments with multi-item post-trip ratings across a large number of real-world trips, this study overcomes several methodological limitations of earlier work and allows a more fine-grained test of competing aggregation heuristics.

The central finding is that retrospective travel satisfaction is best predicted by a Minimum–End heuristic, that is, a combination of the most negative moment during the journey and the final experience. Models based on mean experience, peak–end rules, or duration performed substantially worse. This result suggests that, in public transport contexts, travelers’ remembered evaluations are particularly sensitive to the worst and last moments they encounter. From a theoretical perspective, this finding aligns with work emphasizing negativity bias \citep{rozin2001negativity} and loss aversion \citep{schmidt2002experimental} in evaluative judgments, whereby negative experiences carry disproportionate weight relative to positive ones. It also resonates with research on critical incidents in service encounters, which highlights that singular negative events can dominate overall evaluations even when the remainder of the experience is acceptable or positive \citep{bitner1994critical, friman2001frequency}.

At the same time, the present findings stand in partial contrast to those reported by \citet{abenoza2019does}, who found duration-weighted averaging models to outperform heuristic rules in a multimodal travel context. Several factors may account for this discrepancy. First, the present study relies on withon-trip measurements, whereas \citet{abenoza2019does} used a post-hoc questionnaire where travel satisfaction was rated per trip leg. This difference in measurement granularity may have affected the cognitive processes captured by the respective models. Withon-trip assessments are more likely to reflect momentary experience and local fluctuations, which may favor simpler heuristic integration rules over more effortful averaging processes. In contrast, post-hoc, leg-based evaluations require retrospective integration across segments and time, a context in which duration-weighted averaging may be more cognitively plausible. Furthermore, this study used a multi-item scale, whereas \citet{abenoza2019does}  depended on a single-item measure. Multi-item scales tend to capture a broader and potentially more stable construct of travel experience, whereas single-item measures may be more susceptible to salient moments or recency effects. Taken together, these methodological differences suggest that the relative performance of averaging versus heuristic models may be contingent on when and how travel experience is measured, rather than reflecting a fundamental inconsistency between the two sets of findings.

From a practical perspective, the findings have clear implications for public transport planning and service design. If travelers’ overall evaluations are disproportionately shaped by the worst moments of the journey and by how the trip ends, then interventions that specifically target these critical points may yield substantial gains in remembered satisfaction, even when resources are limited. Reducing the severity or frequency of negative incidents, such as long waits without information, missed connections, or overcrowding during transfers, may be more effective than uniformly improving average service quality. Similarly, ensuring a smooth, predictable, and pleasant final stage of the trip may help “repair” earlier negative experiences. Because remembered satisfaction feeds into attitudes toward public transport and influences future mode choice \citep{lim2024effects}, such targeted measures could contribute to longer-term behavioral change toward sustainable mobility.

Methodologically, the results demonstrate the value of experience sampling approaches for studying travel satisfaction. Measuring travel experience profiles allows researchers to empirically test how different moments contribute to remembered travel satisfaction without imposing strong assumptions about the mapping between objective conditions and subjective perceptions. However, this approach also entails potential limitations. It assumes that travelers are able and willing to accurately report their momentary experiences and that the act of repeated measurement does not substantially alter the experience itself. While prior evidence suggests that momentary measurement generally does not distort overall evaluations \citep{heron2013intensive} 
some studies indicate that frequent reporting may amplify the salience of momentary fluctuations \citep{ariely1998combining}. 

Several limitations of the present study should be acknowledged. Although robustness checks indicate that the observed Minimum–End effects are not solely driven by the last on‑trip questionnaire, future research could administer a follow-up questionnaire a few hours after the journey to further mitigate potential recency effects. Furthermore, although the sample covers multiple cities, the majority of data stem from Hamburg, which may limit generalizability to other transport systems with different service characteristics or user expectations.

Future research could extend this work in several directions. Combining experience sampling with event annotations or contextual sensor data may help identify which concrete incidents drive minimum values in the experience profile. Longitudinal designs could examine how repeated exposure to negative minima shapes stable attitudes toward public transport over time. Finally, integrating real-time experience data into passenger information or feedback systems raises the possibility of adaptive interventions that actively mitigate negative experiences as they occur.

In sum, this study provides evidence that remembered travel satisfaction in public transport is not a simple average of momentary experiences but is shaped by distinct cognitive heuristics, with negative extremes and endings playing a decisive role. Understanding these heuristics offers both theoretical insight into the psychology of travel experience and practical guidance for designing public transport services that leave travelers with better memories, and, ultimately, a greater willingness to choose sustainable modes.

\section{Declaration of generative AI and AI-assisted technologies in the writing process.}

Statement: During the preparation of this work, the authors used ChatGPT Go (Version January 2026) to support spell-checking and refine wording. After using this tool, the authors reviewed and edited the content as needed and take full responsibility for the content of the published article.

\section{Declaration of Interest.}
The authors declare no conflict of interest.

\section{Acknowledgements.}
The authors thank Anke Sauerländer-Biebl for enabling the data collection by providing technical support with data collection. We also thank Jakob Dietze for his support with the data acquisition.

CRediT roles: 
Conceptualization: EB, KI and SB, Data curation, Formal analysis: EB; Funding acquisition: KI; 
Investigation: EB; Methodology: EB and SB; Project administration, Validation, Visualization: EB; Writing – original draft: EB; Writing – review and editing: EB, KI and SB.

Funding: This work was supported by mFUND of the German Federal Ministry of Transportation under the project 'Erlebensatlas' \footnote{https://www.dlr.de/en/ts/research-transfer/projects/erlebensatlas}, grant number 19F1197A, and by the Hamburger Hochbahn AG.

\clearpage
\appendix
\section{High variance model comparisons}
\label{app1}

\begin{table}[ht]
\centering
\caption{Comparison of models predicting post-trip experience variance based on different heuristics. Only participants with a variance of 0.25 (1545 journeys) are included.}
\label{tab:model_comparison_variance}
\vspace{0.3cm}
\begin{tabular}{lccc}
\hline
\textbf{Model} & \textbf{df} & \textbf{AIC} & \textbf{BIC} \\
\hline
Mean        & 4 & 2507.52 & 2523.55 \\
Peakend     & 4 & 2640.93 & 2656.95 \\
MinEnd      & 4 & 2205.49 & 2221.52 \\
End         & 4 & 2350.00 & 2366.03 \\
Min         & 4 & 2556.82 & 2572.85 \\
Duration    & 4 & 3827.97 & 3844.00 \\
\hline
\end{tabular}

\begin{flushleft}
\textit{Note.} df = degrees of freedom (number of estimated model parameters: fixed intercept, fixed slope, random intercept variance, residual variance); AIC = Akaike Information Criterion; BIC = Bayesian Information Criterion. 
Lower AIC/BIC values indicate better model fit.
\end{flushleft}
\end{table}

\begin{table}[ht]
\centering
\caption{Comparison of models predicting post-trip experience variance based on different heuristics. Only participants with a variance of 0.5 (723 journeys) are included.}
\label{tab:model_comparison_variance}
\vspace{0.3cm}
\begin{tabular}{lccc}
\hline
\textbf{Model} & \textbf{df} & \textbf{AIC} & \textbf{BIC} \\
\hline
Mean        & 4 & 1215.40 & 1229.15 \\
Peakend     & 4 & 1324.54 & 1338.29 \\
MinEnd      & 4 & 1068.52 & 1082.27 \\
End         & 4 & 1167.04 & 1180.79 \\
Min         & 4 & 1226.30 & 1240.05 \\
Duration    & 4 & 1854.62 & 1868.37 \\
\hline
\end{tabular}
\end{table}

\begin{table}[ht]
\centering
\caption{Comparison of models predicting post-trip experience variance based on different heuristics. Only participants with a variance of 0.75 (434 journeys) are included.}
\label{tab:model_comparison_variance}
\vspace{0.3cm}
\begin{tabular}{lccc}
\hline
\textbf{Model} & \textbf{df} & \textbf{AIC} & \textbf{BIC} \\
\hline
Mean        & 4 & 767.76 & 779.98 \\
Peakend     & 4 & 814.89 & 827.11 \\
MinEnd      & 4 & 669.00 & 681.22 \\
End         & 4 & 727.68 & 739.90 \\
Min         & 4 & 757.52 & 769.74 \\
Duration    & 4 & 1125.11 & 1137.33 \\
\hline
\end{tabular}
\end{table}

\begin{table}[ht]
\centering
\caption{Comparison of models predicting post-trip experience variance based on different heuristics. Only participants with a variance of 1 (277 journeys) are included.}
\label{tab:model_comparison_variance}
\vspace{0.3cm}
\begin{tabular}{lccc}
\hline
\textbf{Model} & \textbf{df} & \textbf{AIC} & \textbf{BIC} \\
\hline
Mean        & 4 & 457.01 & 467.88 \\
Peakend     & 4 & 495.93 & 506.80 \\
MinEnd      & 4 & 379.75 & 390.62 \\
End         & 4 & 443.94 & 454.81 \\
Min         & 4 & 409.81 & 420.69 \\
Duration    & 4 & 710.75 & 721.63 \\
\hline
\end{tabular}
\end{table}

\clearpage

\section{Comparison with penultimate ontrip travel experience rating}
\label{app2}

\begin{table}[ht]
\centering
\caption{Comparison of models predicting post-trip experience based on different heuristics}
\label{tab:penultimate}
\vspace{0.3cm}
\begin{tabular}{lccc}
\hline
\textbf{Model} & \textbf{df} & \textbf{AIC} & \textbf{BIC} \\
\hline
Mean        & 4 & 3767.97 & 3791.39 \\
Peakend     & 4 & 3920.37 & 3943.79 \\
Peak-penultimate& 4 & 3943.64 & 3967.06 \\

MinEnd      & 4 & 3329.58 & 3352.99 \\
Min-penultimate& 4 & 3448.87 & 3472.29 \\

End & 4 & 3509.11 & 3532.57 \\
Min & 4 & 3857.95 & 3881.37 \\
Penultimate& 4 & 3674.89 & 3698.30 \\

Duration & 4 & 5844.33 & 5867.74 \\
\hline
\end{tabular}

\begin{flushleft}
\textit{Note.} df = degrees of freedom (number of estimated model parameters: fixed intercept, fixed slope, random intercept variance, residual variance); AIC = Akaike Information Criterion; BIC = Bayesian Information Criterion. 
Lower AIC/BIC values indicate better model fit.
\end{flushleft}
\end{table}

\clearpage
\section{8-item-questionnaire only model comparisons}
\label{app3}
\begin{table}[ht]
\centering
\caption{Comparison of models predicting post-trip experience based on different heuristics, using only the 1995 questionnaires with 8 items}
\label{tab:8itemsonly}
\vspace{0.3cm}
\begin{tabular}{lccc}
\hline
\textbf{Model} & \textbf{df} & \textbf{AIC} & \textbf{BIC} \\
\hline
Mean        & 4 & 2414.75 & 2436.50 \\
Peakend     & 4 & 2531.74 & 2553.49 \\
MinEnd      & 4 & 2044.91 & 2066.66 \\
End         & 4 & 2213.93 & 2235.69 \\
Min         & 4 & 2443.96 & 2465.72 \\
Duration    & 4 & 4023.73 & 4045.48 \\
\hline
\end{tabular}

\begin{flushleft}
\textit{Note.} df = degrees of freedom (number of estimated model parameters: fixed intercept, fixed slope, random intercept variance, residual variance); AIC = Akaike Information Criterion; BIC = Bayesian Information Criterion. Lower AIC and BIC values indicate better model fit.
\end{flushleft}
\end{table}

\clearpage
\bibliographystyle{elsarticle-harv} 
\bibliography{02_bibliograohy_sources.bib}

@article{bitner1994critical,
  title={Critical service encounters: The employee's viewpoint},
  author={Bitner, Mary Jo and Booms, Bernard H and Mohr, Lois A},
  journal={Journal of marketing},
  volume={58},
  number={4},
  pages={95--106},
  year={1994},
  publisher={SAGE Publications Sage CA: Los Angeles, CA}
}

@article{friman2004structure,
  title={The structure of affective reactions to critical incidents},
  author={Friman, Margareta},
  journal={Journal of Economic Psychology},
  volume={25},
  number={3},
  pages={331--353},
  year={2004},
  publisher={Elsevier}
}

@article{gremler2004critical,
  title={The critical incident technique in service research},
  author={Gremler, Dwayne D},
  journal={Journal of service research},
  volume={7},
  number={1},
  pages={65--89},
  year={2004},
  publisher={Sage Publications Sage CA: Thousand Oaks, CA}
}

@article{friman2001frequency,
  title={Frequency of negative critical incidents and satisfaction with public transport services. I},
  author={Friman, Margareta and Edvardsson, Bo and G{\"a}rling, Tommy},
  journal={Journal of retailing and consumer services},
  volume={8},
  number={2},
  pages={95--104},
  year={2001},
  publisher={Elsevier}
}

@article{schreiber2000determinants,
  title={Determinants of the remembered utility of aversive sounds.},
  author={Schreiber, Charles A and Kahneman, Daniel},
  journal={Journal of Experimental Psychology: General},
  volume={129},
  number={1},
  pages={27},
  year={2000},
  publisher={American Psychological Association}
}

@article{ariely1998combining,
  title={Combining experiences over time: The effects of duration, intensity changes and on-line measurements on retrospective pain evaluations},
  author={Ariely, Dan},
  journal={Journal of Behavioral Decision Making},
  volume={11},
  number={1},
  pages={19--45},
  year={1998},
  publisher={Wiley Online Library}
}

@article{garling2020review,
  title={Review and assessment of self-reports of travel-related emotional wellbeing},
  author={G{\"a}rling, Tommy and Ettema, Dick and Connolly, Filip Fors and Friman, Margareta and Olsson, Lars E},
  journal={Journal of Transport \& Health},
  volume={17},
  pages={100843},
  year={2020},
  publisher={Elsevier}
}

@article{kahneman1997back,
  title={Back to Bentham? Explorations of experienced utility},
  author={Kahneman, Daniel and Wakker, Peter P and Sarin, Rakesh},
  journal={The quarterly journal of economics},
  volume={112},
  number={2},
  pages={375--406},
  year={1997},
  publisher={MIT Press}
}

@article{van2018influences,
  title={What influences satisfaction and loyalty in public transport? A review of the literature},
  author={Van Lierop, Dea and Badami, Madhav G and El-Geneidy, Ahmed M},
  journal={Transport Reviews},
  volume={38},
  number={1},
  pages={52--72},
  year={2018},
  publisher={Taylor \& Francis}
}

@article{lattman2016perceived,
  title={Perceived accessibility of public transport as a potential indicator of social inclusion},
  author={L{\"a}ttman, Katrin and Friman, Margareta and Olsson, Lars E},
  journal={Social inclusion},
  volume={4},
  number={3},
  pages={36--45},
  year={2016}
}

@article{loewenstein2008hedonic,
  title={Hedonic adaptation and the role of decision and experience utility in public policy},
  author={Loewenstein, George and Ubel, Peter A},
  journal={Journal of Public Economics},
  volume={92},
  number={8-9},
  pages={1795--1810},
  year={2008},
  publisher={Elsevier}
}

@article{raveau2016smartphone,
  title={Smartphone-based survey for real-time and retrospective happiness related to travel and activities},
  author={Raveau, Sebasti{\'a}n and Ghorpade, Ajinkya and Zhao, Fang and Abou-Zeid, Maya and Zegras, Christopher and Ben-Akiva, Moshe},
  journal={Transportation Research Record},
  volume={2566},
  number={1},
  pages={102--110},
  year={2016},
  publisher={SAGE Publications Sage CA: Los Angeles, CA}
}

@article{susilo2014exploring,
  title={Exploring key determinants of travel satisfaction for multi-modal trips by different traveler groups},
  author={Susilo, Yusak O and Cats, Oded},
  journal={Transportation Research Part A: Policy and Practice},
  volume={67},
  pages={366--380},
  year={2014},
  publisher={Elsevier}
}

@article{davelaar2005demise,
  title={The demise of short-term memory revisited: empirical and computational investigations of recency effects.},
  author={Davelaar, Eddy J and Goshen-Gottstein, Yonatan and Ashkenazi, Amir and Haarmann, Henk J and Usher, Marius},
  journal={Psychological review},
  volume={112},
  number={1},
  pages={3},
  year={2005},
  publisher={American Psychological Association}
}

@article{kaiser2020financial,
  title={Financial rewards for long-term environmental protection},
  author={Kaiser, Florian G and Henn, Laura and Marschke, Beatrice},
  journal={Journal of Environmental Psychology},
  volume={68},
  pages={101411},
  year={2020},
  publisher={Elsevier}
}

@article{steinhorst2018effects,
  title={Effects of monetary versus environmental information framing: Implications for long-term pro-environmental behavior and intrinsic motivation},
  author={Steinhorst, Julia and Kl{\"o}ckner, Christian A},
  journal={Environment and Behavior},
  volume={50},
  number={9},
  pages={997--1031},
  year={2018},
  publisher={Sage Publications Sage CA: Los Angeles, CA}
}

@article{de2022attitude,
  title={From attitude to satisfaction: introducing the travel mode choice cycle},
  author={De Vos, Jonas and Singleton, Patrick A and G{\"a}rling, Tommy},
  journal={Transport Reviews},
  volume={42},
  number={2},
  pages={204--221},
  year={2022},
  publisher={Taylor \& Francis}
}

@article{lim2024effects,
  title={Effects of within-trip subjective experiences on travel satisfaction and travel mode choice: A conceptual framework},
  author={Lim, Tommy and Thompson, Jason and Pearson, Lauren and Odgers, Joanne Caldwell and Beck, Ben},
  journal={Transportation Research Part F: Traffic Psychology and Behaviour},
  volume={104},
  pages={201--216},
  year={2024},
  publisher={Elsevier}
}

@article{kahneman1993more,
  title={When more pain is preferred to less: Adding a better end},
  author={Kahneman, Daniel and Fredrickson, Barbara L and Schreiber, Charles A and Redelmeier, Donald A},
  journal={Psychological science},
  volume={4},
  number={6},
  pages={401--405},
  year={1993},
  publisher={SAGE Publications Sage CA: Los Angeles, CA}
}

@article{rozin2001negativity,
  title={Negativity bias, negativity dominance, and contagion},
  author={Rozin, Paul and Royzman, Edward B},
  journal={Personality and social psychology review},
  volume={5},
  number={4},
  pages={296--320},
  year={2001},
  publisher={Sage Publications Sage CA: Los Angeles, CA}
}

@article{schmidt2002experimental,
  title={An experimental test of loss aversion},
  author={Schmidt, Ulrich and Traub, Stefan},
  journal={Journal of risk and Uncertainty},
  volume={25},
  number={3},
  pages={233--249},
  year={2002},
  publisher={Springer}
}

@article{heron2013intensive,
  title={Is intensive measurement of body image reactive? A two-study evaluation using ecological momentary assessment suggests not},
  author={Heron, Kristin E and Smyth, Joshua M},
  journal={Body Image},
  volume={10},
  number={1},
  pages={35--44},
  year={2013},
  publisher={Elsevier}
}

@article{redelmeier1996patients,
  title={Patients' memories of painful medical treatments: Real-time and retrospective evaluations of two minimally invasive procedures},
  author={Redelmeier, Donald A and Kahneman, Daniel},
  journal={pain},
  volume={66},
  number={1},
  pages={3--8},
  year={1996},
  publisher={Elsevier}
}

@article{ariely2000does,
  title={When does duration matter in judgment and decision making?},
  author={Ariely, Dan and Loewenstein, George},
  journal={Journal of experimental psychology: general},
  volume={129},
  number={4},
  pages={508},
  year={2000},
  publisher={American Psychological Association}
}

@article{redelmeier2003memories,
  title={Memories of colonoscopy: a randomized trial},
  author={Redelmeier, Donald A and Katz, Joel and Kahneman, Daniel},
  journal={Pain},
  volume={104},
  number={1-2},
  pages={187--194},
  year={2003},
  publisher={Elsevier}
}

@article{abenoza2019does,
  title={How does travel satisfaction sum up? An exploratory analysis in decomposing the door-to-door experience for multimodal trips},
  author={Abenoza, Roberto F and Cats, Oded and Susilo, Yusak O},
  journal={Transportation},
  volume={46},
  number={5},
  pages={1615--1642},
  year={2019},
  publisher={Springer}
}

@article{bosch2025travel,
  title={Travel experience in public transport: Experience sampling and cardiac activity data for spatial analysis},
  author={Bosch, Esther and Luther, Anna Ricarda and Ihme, Klas},
  journal={Scientific Data},
  volume={12},
  number={1},
  pages={633},
  year={2025},
  publisher={Nature Publishing Group UK London}
}






\end{document}